\documentclass[final,3p,times,twocolumn]{elsarticle}

\usepackage{graphicx}
\usepackage{amssymb}
\usepackage{amsmath}
\usepackage[nodots,nocompress]{numcompress}
\usepackage{bm}
\usepackage{booktabs}
\usepackage{array}

\biboptions{super}
\journal{Solid State Communications}
\begin{document}
\begin{frontmatter}

\title{Nanomechanical AC Susceptometry of an Individual Mesoscopic Ferrimagnet}

\author[1,2] {J.E. Losby}
\author[1,2] {Z. Diao}
\author[1,2] {F. Fani Sani}
\author[1] {D.T. Grandmont}
\author[2] {M. Belov}
\author[1,2] {J.A.J Burgess}
\author[1,2] {W.K. Hiebert}
\author[1,2] {M.R. Freeman}

\address[1]{Department of Physics, University of Alberta, Edmonton, Alberta,
Canada T6G 2G7}

\address[2]{National Institute for Nanotechnology, Edmonton, Alberta,
Canada T6G 2M9}

\ead{mark.freeman@ualberta.ca}

\begin{abstract}
A novel method for simultaneous detection of both DC and time-dependent magnetic signatures in individual mesoscopic structures has emerged from early studies in spin mechanics.  Multifrequency nanomechanical detection of AC susceptibility and its harmonics highlights reversible nonlinearities in the magnetization response of a single yttrium iron garnet (YIG) element, separating them from hysteretic jumps in the DC magnetization. 
\\
\\
\textit{This manuscript was accepted for publication in Solid State Communications (Special Issue: Spin Mechanics).}
\end{abstract}

\end{frontmatter}
\newpage
Nanomechanical torque magnetometry of quasi-static magnetization processes has sparked recent interest due to its exceptional sensitivity (with room-temperature magnetic moment resolution approaching 10$^{6} \mu_{B}$) and ability to non-invasively measure single, mesoscopic elements \cite{moreland, burgess, losby}.  The study of individual structures is a necessity to probe local effects from variation in the magnetic microstructure due to grain boundaries, vacancies, and other intrinsic or extrinsic inhomogeneities.  These are masked in measurements of arrays of magnetic structures but can have a tremendous impact on the magnetization response of single elements.  In this technique a magnetostatic torque, $\bm{\tau}$, is exerted on the magnetization, {\bm{${M}$}}, by an external field, {\bm{$H$}}:  $\bm{\tau}=\bm{M}V \times \mu_0\bm{H}$, in which $V$ is the sample volume.  If the magnetic material is affixed to a torsional resonator, the induced magnetic torque is converted to a mechanical deflection proportional to its magnetization.  For structures with strong in-plane magnetic anisotropy, a small torquing AC (dither) field is applied out-of-plane to the surface while a DC field biases the in-plane magnetization.  By ramping the DC field, the quasi-static magnetization evolution with applied field can be recorded.  

Techniques to probe time-dependent magnetic phenomena yield information complementary to the DC magnetization.  In particular, the AC magnetic response aids in discriminating reversible from irreversible magnetization changes.  In the small signal regime, an AC magnetization $\bm{M^{ac}}=\bm{\chi} \bm{H^{ac}}$ is induced by an alternating field, where $\bm{\chi}$ is a susceptibility tensor.  In bulk magnetic systems, low frequency susceptibility (including the use of higher order susceptibility \cite{rudt,prufer}) measurements are used extensively to monitor myriad phenomena from the onset of ferromagnetism at the Curie temperature \cite{heinrich,sydney}, magnetization reversal processes in thin films \cite{kleeman,stetter} and patterned nanostructures \cite{burgess2}, exchange anisotropy in exchange biased systems \cite{strom}, and dynamics in spin-glass systems \cite{lundgren}.  The advantages offered by nanomechanical transduction have yet to be fully exploited for magnetic susceptibility measurements.  It is highly desirable to develop an experimental technique for measuring both DC and AC components of magnetization, complementary to SQUID measurements \cite{perez, wernsdorfer} by virtue of operating in a wider range of sample environments.  Micro Hall magnetometers can also be very sensitive \cite{novoselov, wirth}, but their low bandwidth will limit applicability to susceptibility measurements.  Synchrotron-based XMCD-PEEM has also been utilized to record low-frequency dynamic susceptibility in magnetic thin-films \cite{aspelmeier,romer}.

Here we report a nanomechanical platform for simultaneous DC and AC magnetic measurements through the introduction of an AC field component along the bias field direction, to probe the quasi-static longitudinal susceptibility.  Two orthogonal AC fields, if applied at different frequencies simultaneously, give rise to AC magnetic torques at the sum and difference frequencies.  These can be tuned to match the natural resonance frequency of the resonator, allowing it to function effectively as a signal mixer.  Different frequency components can be extracted from the mechanical signal as the DC field sweeps the magnetization texture to measure simultaneously the DC magnetization, AC susceptibility, and higher order AC susceptibility terms. 

\begin{figure}[h]
\includegraphics[scale=1]{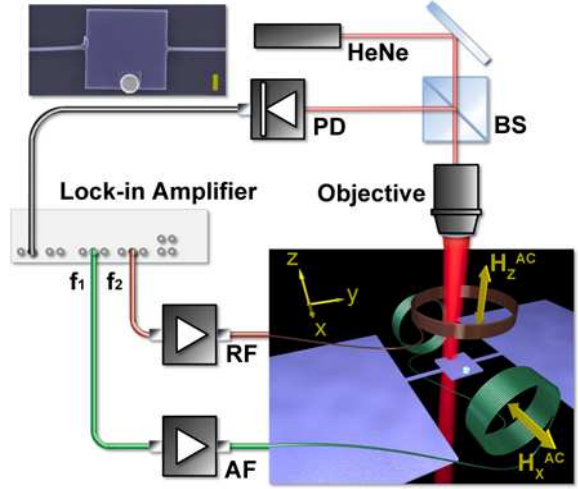}
\caption{Instrument schematic for frequency-mixed nanomechanical detection of AC susceptibility.  The lock-in amplifier provides the reference/drive frequencies, $f_{1}$ and $f_{2}$, which induce orthogonal AC fields through a Helmholtz coil assembly ($H_{x}^{ac}$) and single coil ($H_{z}^{ac}$), both illustrated in the bottom-right inset).  The lock-in output signal is amplified by audio (AF) and radio (RF) frequency power amplifiers, respectively.  A HeNe laser is used to interferometrically detect the mechanical motion of the torsional resonator (BS=beam splitter, PD=photodetector).  The difference and/or sum frequencies are demodulated simultaneously at the lock-in.  Top left inset: false-colour scanning electron micrograph of the device (scale bar = 1 $\mu$m)}
 \label{fig:vel}
 \end{figure} 

Determination of the susceptibility requires some level of description of the spin-mechanical coupling, taking in to account the conversion of magnetic torque into mechanical torque.  The full tensor analysis will include Einstein-de Haas terms, which we will neglect here to concentrate on the net mechanical torque arising through magnetic anisotropy.  Conceptually, we can view the system as two torsion springs connected in series.  The geometrically-confined magnetization texture of the disk is the first spring, as is the one acted on directly by the torquing fields.  In turn, the net magnetic torque will twist the torsion rod until an equivalent mechanical counter-torque develops.  For the present analysis, we define the effective magnetic torsion spring constants by converting magnetic susceptibilities into angular changes as if the magnetization of the element were a single macrospin.  Although far from the real case, the parallel axis theorem allows for the same net torque from the same net magnetization independent of how the magnetization is distributed. 

The net magnetic anisotropy (shape plus magnetocrystalline) of the mesoscale magnetization in this simple picture manifests itself through differences in diagonal components of the magnetic susceptibility tensor.  If no anisotropy exists, then the equilibrium magnetization will always be parallel to the applied field, with a resultant torque of zero.  In general, one will encounter structures with significantly different susceptibilities parallel and perpendicular to the plane of the torsion paddle, but where neither component is negligible.  This is the case for the specific example we consider below, a 3D vortex state in a short cylinder of YIG, where the low-field linear in-plane susceptibility is approximately twice the out-of-plane susceptibility \cite {losby2}.  Then, the resultant torque in simultaneous small (for linear magnetization response) fields $H_{x}$ and $H_{z}$ is:
\begin{equation}
-\tau_y = \mu_0(M_{x}H_{z}-M_{z}H_{x})V= \mu_0(\chi_{x}-\chi_{z})H_{x}H_{z}V
\end{equation}
This net torque will transfer to the lattice and twist the resonator.  For a low frequency AC field, an AC torque ensues in phase with the field.  If the frequency is not too high, the magnetization remains in quasi-static equilibrium with the field at all times.  For the vortex structure in the absence of pinning (no slow thermal dynamics), this criterion is satisfied at frequencies in the low MHz regime, convenient for resonant enhancement of the mechanical response in nanoscale torsional structures.  In the present experiment, corrections for the applied field angles changing in the frame of reference of the paddle are negligible: the mechanical spring is much stiffer ($\sim 10^6 \times$) than the magnetic spring, and the mechanical Q is only 2600.

When two orthogonal AC fields act simultaneously at different frequencies, $f_{1}$ and $f_{2}$, AC magnetic torques at the sum and difference frequencies also arise, according to:
\begin{equation}
\begin{split}
-\tau_{y} = \mu_0(\chi_{x}-\chi_{z})H_{x}^{ac}\cos (2\pi f_{1}t)H_{z}^{ac}\cos (2\pi f_{2}t)V \\
= \mu_0(\chi_{x}-\chi_{z})\frac{H_{x}^{ac}H_{z}^{ac}}{2}(cos (2\pi(f_{2}-f_{1}) t ) + cos (2\pi(f_{2} + f_{1}) t) )V
\end{split}
\end{equation}
With the simultaneous application of a DC field to sweep the magnetization texture through a hysteresis cycle, this becomes the basis of the AC susceptometry measurement, to complement the DC magnetization data. 
 
 \begin{figure}[ht]
\includegraphics[scale=1]{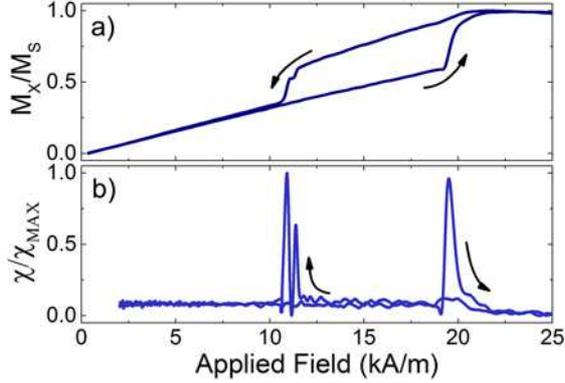}
\caption{DC torque magnetometry of an individual micromagnetic YIG disk, (a) and the corresponding numerical derivative, (b).}
 \label{fig:vel}
 \end{figure}

For demonstration of the technique, a single-crystal, mesoscale YIG disk (radius = 600 nm, thickness = 500 nm) was focused ion beam (FIB)-milled from an epitaxial thick film and nanomanipulated \textit{in-situ} onto a prefabricated torsional resonator \cite{losby2}.  A scanning electron micrograph of the completed device is shown in the inset of Fig 1.  The FIB milling and `pick and place' procedures were employed to overcome the difficulty of forming monocrystalline mesoscale objects which are to be measured singularly.  The pristine magnetic nature of the YIG disk is exhibited, within experimental resolution, by the lack of Barkhausen signatures in the change of the magnetization with applied field, which are associated with nanoscale imperfections such as grain boundaries in polycrystalline materials.  In the present context YIG simplifies interpretation of the demonstration susceptibility signals.  Hysteretic minor loops associated with Barkhausen transitions in the magnetization evolution \cite{burgess}, can strongly influence and complicate the AC susceptibility. The YIG disk, on the other hand, was shown to have no observable minor hysteresis.

The resonator was driven by the magnetic torque exerted on the YIG disk.  The mechanical deflection was optically detected through the interferometric modulation of the light reflected from the device.  The magnetization was biased by a variable external field, ($H_{x}^{dc}$, Cartesian coordinates indicated in Fig. 1) in the plane of the resonator and perpendicular to the torsion rod.  The AC torquing field was supplied by a small copper wire coil which produced an out-of-plane dither field, $H_{z}^{ac}$, to complete the field orientations necessary for the torque magnetometry measurement.  The AC field was driven at the fundamental torsion frequency to maximize detection sensitivity.  The magnetization of the YIG disk with applied bias field is shown in Fig. 2,  characterized by a unipolar hysteresis similar to what has been observed in two-dimensional disks where the sharp transitions are attributed to the nucleation and annihilation of a magnetic vortex.  However, the thickness of the YIG disk promotes additional three dimensional structure in the magnetization texture (detailed in Ref. 19).

A result from the simultaneous acquisition of the DC magnetization and AC susceptibility is shown in Fig 3.  Features agreeing closely to the numerical derivative of the DC magnetization curve, Fig. 2b, are evident with the differences arising from the susceptibility screening out irreversible events in the field sweep.  Here, the lock-in amplifier provided both orthogonal ac drive frequencies, $f_{1}$ ($H_{x}^{ac}$) and $f_{2}$ ($H_{z}^{ac}$), while recording the response of the resonator through simultaneous demodulation \cite{zurich} at $f_{2}$ and at $f_{2} - f_{1}$ (see Fig. 1 for measurement scheme).  It is interesting to note that a softening of the magnetization texture (peak in the susceptibility) occurs just after the irreversible annihilation transition on the field sweep-up.  
The highest field susceptibility is at opposite phase in this measurement consistent with the possibility inherent in Eqn. 2 for $\chi_z$ contributes to exceed the $\chi_x$ contribution as the magnetization approaches saturation in $x$. 
This feature is replicated again on the sweep down at a slightly lower field -- a hysteretic feature not readily apparent from the DC magnetization.

  \begin{figure}[h]
\includegraphics[scale=1]{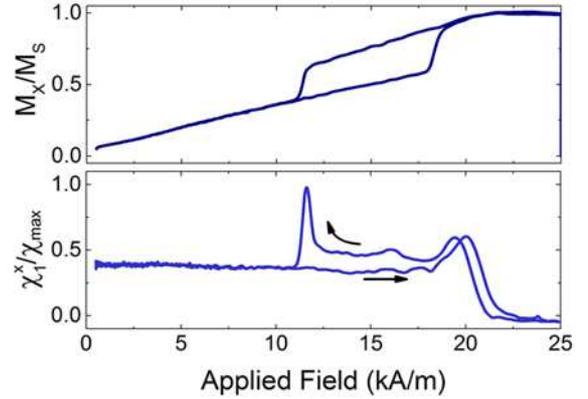}
\caption{Simultaneous acquisition of the DC magnetometry through the driven signal at $f_{2}$=$f_{res}$+$f_{1}$, a), and low frequency AC susceptibility detected at the mixing (resonance) frequency, $f_{res}$, b). Here, $f_{1}$ was set to 500 Hz.}
 \label{fig:vel}
 \end{figure}

The technique can easily be extended to detect the harmonics of AC susceptibility arising when the magnetization response is nonlinear in field.  For example, for a magnetization describable by a polynomial series in field, 
 \begin{equation}  
M_x = a_1H_x +a_2H^{2}_{x}+a_3H^{3}_{x}+a_4H^{4}_{x}+a_5H^{5}_{x}+a_6H^{6}_{x}+\dots,
\end{equation}

where  $H_x = H_x^{dc}+H_x^{ac}$, multiple harmonics of $f_{1}$ arise and also mix with $f_{2}$ to generate unique torque terms.  For this representation of nonlinear magnetization, Table 1 shows the amplitude coefficients for the first six entries in the Fourier sum for the full nonlinear susceptibility,
\begin{equation}
 \chi_{x}(H_x) = \sum_{n=-\infty}^{\infty} \chi^{x}_{n}(H_x)e^{-i(2\pi nf_1) t} . 
\end{equation}

 \begin{table}[h]
	\label{t1}
  \caption{Harmonics of the AC susceptibility}
  \begin{tabular}{ l  || p{4cm}  || l  }
    \hline
    $f_2\pm nf_1$ & Amplitude coefficients  & related to   \\ \hline
    
    $f_2 + 0 $  & $a_1H_x^{dc}+a_2[(H_x^{dc})^2 + \frac{1}{2}]+a_3[(H_x^{dc})^3 + \frac{3}{2}H_x^{dc}]   $    
    
   $  +a_4[(H_x^{dc})^4+3H_x^{dc})^2+\frac{3}{8}]+a_5[(H_x^{dc})^5+5(H_x^{dc})^3+\frac{15}{8}H_x^{dc}]$
     
     $ +a_6[(H_x^{dc})^6 + \frac{15}{2}(H_x^{dc})^4 + \frac{45}{8}(H_x^{dc})^2 + \frac{5}{16}]$  & $\chi^{x}_{0}(H_x)$ \\ \hline
		
     $f_2 \pm f_1 $ & $a_1 +2a_2H_x^{dc}  + a_3[3(H_x^{dc})^2 + \frac{3}{4}]+ a_4[4(H_x^{dc})^3 + 3H_x^{dc}] $
          
   $  + a_5[5(H_x^{dc})^4 + \frac{15}{2}(H_x^{dc})^2 + \frac{5}{8}]  $	
   
   $+a_6[6(H_x^{dc})^5 + 15(H_x^{dc})^3 + \frac{15}{4}H_x^{dc}] -\ \frac{dM_z}{dH_z}|_{H_x^{dc}}$ & $\chi^{x}_{1}(H_x)$ \\  \hline
		
		 $f_2 \pm 2f_1 $  & $ \frac{1}{2}a_2 + \frac{3}{2}a_3H_x^{dc} + a_4[3(H_x^{dc})^2 + \frac{1}{2}] +a_5[5(H_x^{dc})^3 + \frac{5}{2}H_x^{dc}]$
		 
		 $+a_6[ \frac{15}{2} (H_x^{dc})^4 + \frac{15}{2}(H_x^{dc})^2 + \frac{15}{32}]$  & $\chi^{x}_{2}(H_x)$\\  \hline
		
		 $f_2 \pm 3f_1$  & $ \frac{1}{4}a_3 + a_4H_x^{dc} + a_5[\frac{5}{2}(H_x^{dc})^2 + \frac{5}{16}] + a_6[5(H_x^{dc})^3 + \frac{15}{8}H_x^{dc}] $ & $\chi^{x}_{3}(H_x)$\\ \hline
		
		 $f_2 \pm 4f_1$  & $ \frac{1}{8}a_4 + \frac{5}{8}a_5H_x^{dc} + a_6[\frac{15}{8}(H_x^{dc})^2 + \frac{3}{16}]$  & $\chi^{x}_{4}(H_x)$\\ \hline
		
		 $f_2 \pm 5f_1$  & $ \frac{1}{16}a_5 + \frac{3}{8}a_6H_x^{dc} $ & $\chi^{x}_{5}(H_x)$ \\ \hline
		
		 $f_2 \pm 6f_1$  & $\frac{1}{32}a_6 $  & $\chi^{x}_{6}(H_x)$\\ 
		
    \hline
  \end{tabular}
	\end{table}
 
A spectroscopic frequency-versus-field mapping of the higher order AC susceptibilities for the YIG disk is shown in Fig. 4 for the sweep-down portion of the hysteresis cycle (31 to 4 kA/m).  Here the frequency response was recorded at each magnetic field step while keeping both $f_1$ ($H_x^{ac}$) and $f_2$ ($H_z^{ac}$) constant (500 Hz and 1.8095 MHz, respectively). The $f_2$ drive was shifted from the resonance frequency (1.808 MHz) in order to allow for more susceptibility harmonics to be recorded within the mechanical resonance line-width.  The brightest band is the demodulation of the $f_2$ drive, while the subsequent bands represent the harmonics of the magnetic susceptibility.  The features observed in the second half of the field sweep in Fig. 3b are reproduced, including the strong susceptibility peak just before (very close to the onset of) the nucleation transition.  Through such strong nonlinearities in the magnetization curve, mixing frequencies $f_2 \pm nf_1$ are observed up to $n=7$ in Fig. 4.  

\begin{figure}[h]
\includegraphics[scale=1]{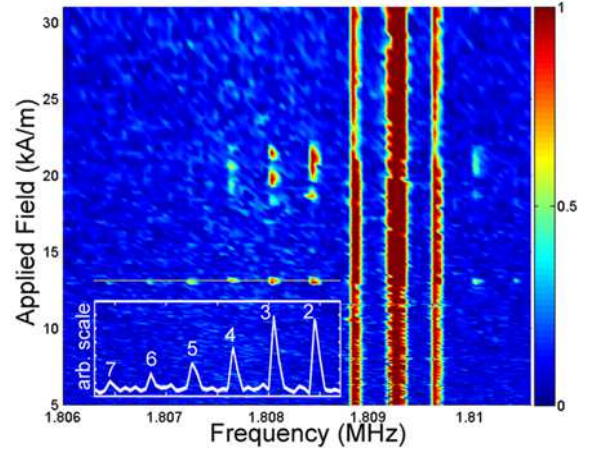}
\caption{Spectroscopic mapping of higher harmonics of the AC susceptibility.  $f_1$ ($H_x^{ac}$) and $f_2$ ($H_z^{ac}$) were each driven simultaneously at constant frequency (500 Hz and 1.8095 MHz respectively) while the frequency response (100 Hz bandwidth) was measured with magnetic field.  The image shows the sweep-down portion of the field sweep.  The band of highest intensity is the demodulation of $f_2$, with the successive bands representing the harmonics of AC susceptibility, $f_2 \pm n f_1$.  A line-scan thorough the high susceptibility peaks is shown in the bottom left, showing up to the n=7 harmonic.  The original intensity scale was cropped to show the highest order signals.}
 \label{fig:vel}
 \end{figure}
 
The multifrequency responses are all harmonics of the low AC frequency
dithering the magnetization, mixed to fall within the mechanical
resonance linewidth. ÊHigh harmonics of susceptibility have been used
previously to characterize hysteresis for bulk ferromagnetism and
superconductivity. ÊIn bulk systems, and in arrays of nanostructures,
the strongest nonlinearities in magnetization come from hysteretic
features. ÊTypical AC field amplitudes can probe minor hysteresis
loops \cite{bean}, or even the full hysteresis close enough to the
Curie temperature \cite{rudt}. ÊBy contrast, the single mesoscale
structure we examine here exhibits no minor hysteresis, and its
magnetization nonlinearities are observed without any diminution from
array averaging. ÊThe harmonics characterize the non-hysteretic, nonlinear
response.

In summary, nanomechanical detection of the low-frequency AC susceptibility (along with higher order harmonics) of an individual micromagnetic disk was accomplished through frequency mixing of orthogonal 
AC driving fields manifesting as a mechanical torque on the resonator.  In addition to providing further insight into quasi-static magnetization processes in geometrically-confined magnetic elements, this signal-mixing approach, in principle, will enable very broadband measurements of frequency-dependent susceptibility.
\section*{Acknowledgements}
The authors would like to thank NSERC, CIFAR, Alberta Innovates and NINT for support.  Partial device fabrication was done at the University of Alberta NanoFab.  FIB milling and manipulation was carried out by D. Vick and S. R. Compton at the NINT electron microscopy facility.

\end{document}